\documentclass[reprint,onecolumn,english,nofootinbib,letterpaper,nobalancelastpage]{revtex4-1}

\usepackage[utf8x]{inputenc}
\usepackage{amsmath}
\usepackage{amssymb}
\usepackage{bbm}
\usepackage[english]{babel}
\usepackage{graphicx}
\usepackage{verbatim}
\usepackage{latexsym}
\usepackage{mathptm}
\usepackage{color}

\newcommand{\f}{\begin{equation}}
\newcommand{\ff}{\end{equation}}

\begin{document}
\title{The DSR-deformed relativistic symmetries and the relative locality of 3D quantum gravity}

\author{$~$\\
{\bf Giovanni AMELINO-CAMELIA}$^{1,2}$,
  {\bf Michele ARZANO}$^{1,2}$, {\bf Stefano BIANCO}$^{1,2}$, {\bf Riccardo J.~BUONOCORE}$^{1}$
\\
$^1${\footnotesize Dipartimento di Fisica, Sapienza University of Rome, P.le Moro 2, 00185 Roma, Italy }\\
$^2${\footnotesize Sez.~Roma1 INFN, P.le Moro 2, 00185 Roma, Italy }}

\begin{abstract}
Over the last decade there were significant advances in the understanding of quantum gravity
coupled to point particles in 3D (2+1-dimensional) spacetime.
Most notably it is emerging that the theory can be effectively described
as a theory of free particles on a momentum space with anti-deSitter geometry and with noncommutative spacetime coordinates of the type $[x^{\mu},x^{\nu}]=i \hbar \ell \varepsilon^{\mu\nu}_{\phantom{\mu\nu}\rho} x^{\rho}$.
We here show  that the recently proposed relative-locality curved-momentum-space framework is ideally suited for accommodating these structures characteristic of 3D quantum gravity.
Through this we obtain an intuitive characterization of the DSR-deformed Poincar\'e symmetries of 3D quantum gravity, and find that the associated
relative spacetime locality is of the type producing dual-gravity lensing.
\end{abstract}

\maketitle





\section{Introduction}\label{secintro}
We here report a study which is relevant for two of the most active areas of quantum-gravity research over the last decade.
Some aspects
of our analysis contribute to the ongoing development
 of DSR-deformed relativistic symmetries at the Planck scale,
while other aspects of our analysis are inspired
by previous studies of quantum-gravity in 3D (2+1-dimensional) spacetime.

The first studies of DSR-deformed relativistic symmetries
intended~\cite{dsr1,dsr2}
to provide an alternative
interpretation of results on quantum-gravity modifications of special relativistic laws,
such as modifications of the on-shell relation of the
type $p^2 = E^2 - m^2 + \Delta_{QG}(\ell ,E)$, with  $\Delta_{QG}(\ell ,E)$ some quantum-gravity correction and $\ell$ expected to be given roughly by the inverse of the Planck scale.
At first this quantum-gravity-research results producing laws not compatible with special
relativity were interpreted
as inevitably associated with the presence of a non-relativistic preferred-frame picture.
Starting with
Ref.~\cite{dsr1,dsr2} (and now finding support in a rather sizable literature,
see, {\it e.g.}, Refs.~\cite{jurekDSR1,leeDSRprd,jurekDSR2,leeDSRrainbow,gacdsrREVIEW2010})
it was understood that some of these modifications of special-relativistic laws could
be accommodated in scenarios (``DSR scenarios") that are still fully relativistic, preserving the
principle of equivalence of inertial frames, if  one allows for
 $\ell$-deformed laws of transformation  between observers~\cite{dsr1,dsr2}.
These relativistic proposals then have as invariant characteristic scales of the transformation rules
not only the speed-of-light scale $c$ (here mute because of our choice of units $c=1$)
but also the Planck scale $\ell^{-1}$.

Interest in the 3D quantum-gravity problem started to pick up during the
1980s~\cite{Deser:1983tn, Deser:1988qn, Witten:1988hc, Witten:1989sx}.
Our study is primarily connected to more recent work
coupling 3D gravity to point
particles~\cite{Bais:1998yn,Bais:2002ye,Meusburger:2003ta,Meusburger:2005in,Schroers:2011wn,Osei:2011ig,Freidel:2005me,Freidel:2005bb},
showing that several potentially different approaches agree on some results, which at this point
should then be viewed as robust. In particular, it is found
that the momenta of the particles are described by elements of the isometry group of the ``model space-time" which provides gluing data for the non-trivial topology describing them.  The first intuition of this can be found in studies from the
1990s~\cite{'tHooft:1993nj,'tHooft:1996uc,Matschull:1997du} in a metric formalism.
More recent refined descriptions~\cite{Bais:1998yn,Bais:2002ye,Meusburger:2003ta,Meusburger:2005in,Schroers:2011wn,Osei:2011ig}
established results such
that the momentum of particles coupled to Chern-Simons gravity is given by holonomies of the gauge group of the theory along non-contractible loops containing the puncture describing the particle.  The connection between metric descriptions  and Chern-Simons descriptions
was investigated in Refs.~\cite{Meusburger:2003ta,Meusburger:2005in}.
For reasons that shall be clarified also by our line of analysis (see later) the momentum-space
features of this characterization can be deduced already at the level of the classical
theory, and they persist when the theory is
quantized~\cite{Freidel:2005me,Freidel:2005bb,Osei:2011ig,Schroers:2011wn}.
And for the quantum theory the counterpart of this non-trivial geometry of momentum space turns out
to be noncommutativity of the spacetime coordinates.

We are interested in the case of 3D gravity without a cosmological constant, where
one ends up with a momentum space with anti-deSitter geometry, and noncommutativity
of the spacetime coordinates of the type~\cite{majidSPINNING}
\begin{equation}
[x^{\mu},x^{\nu}]=i \hbar \ell \varepsilon^{\mu\nu}_{\phantom{\mu\nu}\rho} x^{\rho}
\label{spinST}
\end{equation}
which is the case we here label\footnote{This is inspired by the analogy between (\ref{spinST})
and the angular-momentum algebra (which could also suggest the name ``spin spacetime"~\cite{majidSPINNING}).} ``spinning noncommutative spacetime".

Crucial for our analysis is the observation that these features of curvature of momentum
space and noncommutativity of spacetime coordinates have already provided the starting point
for some DSR-relativistic scenarios. In particular, there was a considerable amount
of work on a DSR scenario centered on~\cite{flaviogiuliaDESITTER,anatomy,goldenrule}
a momentum space with de Sitter geometry
and spacetime noncommutativity of $\kappa$-Minkowski\footnote{Interestingly,
also the possibility of $\kappa$-Minkowski noncommutativity can arise in the 3D context of Chern-Simons
theories, but only at the cost of renouncing~\cite{meusbschrKAPPA} to some aspects of
the relevance for the Einstein-Hilbert action.}
   type~\cite{majrue,lukieANNALS}.
We here show that the techniques and approaches developed in those contexts can indeed be adapted to
the scenario inspired by the 3D quantum gravity. In particular, the ``relative-locality curved-momentum-space
framework"~\cite{prl,grf2nd}, which had been valuably applied to
the $\kappa$-Minkowski-based picture~\cite{flaviogiuliaDESITTER,anatomy},
is here found to be also applicable to the 3D-gravity-inspired picture.
The relative-locality framework can be applied to an even wider class of theories,
but specifically in the context of DSR-relativistic theories it empowers us to properly
implement within a spacetime picture 
the deformations of  translation transformations that are typically encountered.
This will also play a key role in our analysis.

The most significant results we obtain establish that, as preliminarily suggested
by some previous studies (see, {\it e.g.}, Refs.~\cite{Freidel:2005me,Freidel:2005bb,kodadsr,kodadsrJUREK,oriti3D}),
3D quantum gravity is a theory with DSR-deformed relativistic symmetries.
And we show that a characteristic aspect of our 3D-gravity-inspired analysis
is ``dual-gravity lensing", one of the least studied among possible features
of a scenario with deformation of relativistic symmetries (previously considered explicitly
only in Refs.~\cite{leelaurentGRB,transverse}).

We feel that there are rather general benefits in performing studies such as ours 
at the interface between research on DSR-deformed relativistic symmetries and 
research on 3D quantum gravity.
On the DSR side one should notice that the construction of 4D models with
DSR-deformed relativistic symmetries is at present at an advanced but incomplete stage,
and the well-understood 3D-quantum-gravity context can be ideally suited for giving guidance toward
uncovering other significant implications of these deformations.
The debate on DSR often revolves around whether these relativistic deformations should at all
be considered in relation to the quantum-gravity problem, and the fact that they necessarily arise
in the 3D-quantum-gravity context surely provides a strong element of support for advocates
of the study of DSR-deformed relativistic symmetries.
And the well-understood 3D-quantum-gravity context is also
 ideally suited for giving guidance on the conceptual side: features like relative locality and noncommutativity
 of momentum-compositions laws may appear puzzling when introduced by hand in a DSR picture, so the fact
 that we expose here their inevitability in the 3D-gravity context can change the balance of intuitions
 on such features.
 Moreover, 3D quantum gravity also provides  an explicit example of the
 sort of mechanisms which are expected to produce DSR-deformed laws of kinematics:
for 3D quantum gravity we can actually integrate out gravity~\cite{Freidel:2005me,Freidel:2005bb}
   and verify that its effects are reabsorbed into novel relativistic
   properties for the gravity-free propagation of particles.
It is not unnatural to conjecture that also in 4D quantum gravity
there would be some regime of observation such
that the only  quantum-gravity effects there tangible can be reabsorbed
into novel relativistic  properties for the gravity-free propagation of particles,
but in the 4D case providing explicit examples where this intuition applies is beyond
our present technical abilities.

On the 3D-gravity side we stress how reliance on expertise gained in previous studies
of DSR deformations might amplify the potentialities for 3D results to inspire
phenomenological programmes for real 4D quantum gravity. Surely 4D quantum gravity will be very different
from its 3D version, but it is legitimate to speculate that some of the features uncovered in the
much simpler 3D context might also apply to the 4D context we are really interested in.
But such a legitimate speculation could be valuable only if it can be scrutinized experimentally,
whereas most results on 3D quantum gravity so far have been of rather formal nature. By uncovering a
role for DSR-deformed relativistic symmetries in the 3D-gravity context we here open the way
for using 3D-gravity as guidance for proposals of ``DSR phenomenology", some of which are
already at rather
advanced stage of development (see, {\it e.g.}, Refs.~\cite{sethdsr,dsrphen,unoEdue}).

As mentioned, we adopt units such that the speed-of-light scale
is set to $1$, and
we denote by $\ell$ the
inverse of
3D-quantum-gravity Planck scale.
It turns out to be sufficient for our purposes to assume $\ell$ is very small,
and therefore several of our results are shown  only at leading (linear) order
in $\ell$.
Also note that for the antisymmetric tensor  $\varepsilon_{\mu\nu\rho}$
we adopt conventions such that $\varepsilon_{012}=-1$ and indices are raised and lowered with the metric $\eta_{\mu\nu}=(-1, 1, 1)$. This in particular implies that the
defining commutation relations for the spinning spacetime
could also be written
as $[x^1,x^2]=-i \hbar \ell x^0$, $[x^1,x^0]=-i \hbar \ell x^2$, $[x^2,x^0]=i \hbar \ell x^1$.

\section{Anti-de Sitter momentum space and spinning spacetime}\label{secgeometry}
Before getting started with our analysis it useful for our purposes to characterize quantitatively, in this section,
the known facts about 3D gravity and quantum gravity that were described qualitatively in our opening remarks.
None of the points
made in this section is original, since they can all be found here and there
in the literature (see in particular Refs.~\cite{Matschull:1997du,Freidel:2005me}),
but there is a character of originality in the content of this section since no previous study had
looked at these results from a perspective like ours. Therefore the organization
of known results we here give cannot be found in any single previous publication.

We focus on the case without a cosmological constant, and we are mainly interested in the
connection between geometry of momentum space and spacetime noncommutativity.
We already mentioned that the momentum space has anti-de-Sitter geometry, but more precisely
the momentum space is the Lie group $SL(2,R)$,
 group of linear transformations acting on $\mathbb{R}^2$, with determinant equal to one. And our first task
 is to expose the anti-de-Sitter geometry of this momentum space.
 To do this, we note that it is possible to write the generic element ${\bf p}$ of $SL(2,R)$ as a combination of the identity
matrix and of the elements of a basis of $sl(2,R)$, the Lie algebra of $SL(2,R)$:
\begin{equation}
 {\bf p} = u\mathbb{I}-2\xi_{\mu}X^{\mu}.
\label{matrixp}
\end{equation}
Here $\mathbb{I}$ is the identity $2\times 2$ matrix and the $X^{\mu}$ are
\begin{equation}
X^0=\frac{1}{2}
\begin{pmatrix}
    0 & 1 \\
   -1 & 0 \\
\end{pmatrix},
X^1 =\frac{1}{2}
\begin{pmatrix}
  0 & -1 \\
  -1  & 0 \\
\end{pmatrix},
X^2 =\frac{1}{2}
\begin{pmatrix}
   -1 & 0 \\
   0 & 1 \\
\end{pmatrix},
\end{equation}
which constitute a basis of $sl(2,R)$, and the requirement of having determinant equal to one
($\det {\bf p}=1$) implies that the parameters $u,\xi_{\mu}$ must be constrained to satisfy
\begin{equation}
 u^2-\xi^{\mu}\xi_{\mu}=1~.
\label{adS}
\end{equation}
This constraint provides, as announced,
 the definition of a 3 dimensional anti-de Sitter geometry.

Among the choices of coordinates for this momentum-space geometry used in the 3D-gravity
literature,
 particularly convenient for our purposes
is the choice of coordinates
of coordinates $p^{\mu}$ such that\footnote{An alternative coordinatization which is also
frequently used (see, {\it e.g.}, Ref.~\cite{Matschull:1997du})
adopts coordinates $P_\alpha , P_\beta , P_\gamma$ which are essentially Euler angles
 and are related to the coordinates of our Eq.~(\ref{ourcoord}) by the relations
$ p_0=\ell^{-1}\sin(P_\alpha\ell) \cosh(P_\beta\ell)$, $ p_1=\ell^{-1} \cos(P_\gamma\ell) \sinh (P_\beta\ell)$,
$ p_2=\ell^{-1} \sin(P_\gamma\ell) \sinh (P_\beta\ell)$.}
\begin{equation}
 {\bf p}= \sqrt{1+\ell^2p^{\mu}p_{\mu}}\mathbb{I}-2\ell p_{\mu}X^{\mu}~,
\label{ourcoord}
\end{equation}
since we shall find that this choice of coordinates allows one
to describe the metric very compactly and
to formulate
the law of composition of momenta as very explicitly given in terms of
the algebraic properties of the $X^{\mu}$ matrices.

The on-shell condition satisfied by these momenta can be
derived~\cite{Matschull:1997du,Freidel:2005me,majidSPINNING}
exploiting the fact that our momentum space is a Lie group, and it
is therefore possible to define a metric over this space using the Killing form of its Lie algebra.
This leads to the result~\cite{Matschull:1997du,Freidel:2005me,majidSPINNING}
\begin{equation}
 \ell^{-2}\left(\arcsin\left(\sqrt{-\ell^2p^{\mu}p_{\mu}}\right)\right)^2=m^2~.
 \label{jocnn}
\end{equation}
We shall rederive this result from a somewhat different perspective in Sec.~IV.

Next we must notice that the the group structure of our momentum space implies
that the law of composition of momenta is nonlinear.
In fact, if we multiply two group elements
\begin{equation}
\begin{array}{l}
{\bf p}=\sqrt{1+\ell^2p^{\mu}p_{\mu}}\mathbb{I}-2\ell p^{\mu}X_{\mu}\\
{\bf q}=\sqrt{1+\ell^2q^{\mu}q_{\mu}}\mathbb{I}-2\ell q^{\mu}X_{\mu}
\end{array}
\end{equation}
we obtain a new element ${\bf pq}$, and there is a simple but nonlinear relation
between the coordinates $(p \oplus q)_{\mu}$ of ${\bf pq}$ and the
coordinates $p_{\mu} , q_{\mu}$ of ${\bf p}$,${\bf q}$:
\begin{equation}
\label{somma}
(p \oplus q)_{\mu} = \sqrt{1+\ell^2 q^{\mu}q_{\mu}}p_{\mu}+\sqrt{1+\ell^2 p^{\mu}p_{\mu}} q_{\mu}-\ell \varepsilon_{\mu}^{\phantom{\mu}\nu\rho}p_{\nu}q_{\rho}~,
\end{equation}
which we derived using the identity
\begin{equation}
 \label{prodotto-di-X}
X^{\mu}X^{\nu}=\frac{1}{4}\eta^{\mu\nu}\mathbb{I}+\frac{1}{2}\varepsilon^{\mu\nu}_{\phantom{\mu\nu}\rho}X^{\rho}~.
\end{equation}

Finally, we notice that (\ref{prodotto-di-X}) implies that the $X^{\mu}$
satisfy by construction (up to a dimensionful constant)
the commutation relations of the spinning spacetime
\begin{equation}
 \label{finallyspinning}
[X^{\mu},X^{\nu}]=\varepsilon^{\mu\nu}_{\phantom{\mu\nu}\rho}X^{\rho}~.
\end{equation}
So our perspective on the implications of the fact that the relevant momentum space is a Lie group
also hints at the role that the spinning spacetime plays
(as indicated in various ways by the 3D-gravity literature)
in 3D quantum gravity.

\section{Relativistic kinematics on the spinning spacetime}\label{secdsr}
Having devoted the previous section to known
 3D-quantum-gravity results, summarized adopting a perspective which prepares our analysis,
we are now ready for performing the first task of our analysis, which is the one of exposing the
DSR-relativistic symmetries of the emerging framework.
We shall be satisfied analyzing the classical limit of the construction described in the previous section,
characterized by spacetime coordinates with Poisson brackets given by
\begin{equation}
 \label{pncc}
 \{x^{\mu},x^{\nu}\}=\ell \varepsilon^{\mu\nu}_{\phantom{\mu\nu}\rho} x^{\rho}~,
\end{equation}
and by a momentum space with coordinates $p_{\mu}$ constrained on
 mass shells governed by
\begin{equation}
\label{mass-shellDOPPIONE}
 \ell^{-2}\left(\arcsin\left(\sqrt{-\ell^2p^{\mu}p_{\mu}}\right)\right)^2=m^2~,
\end{equation}
and with law of composition
\begin{equation}
\label{composizione}
(p \oplus q)_{\mu} = \sqrt{1+\ell^2 q^{\mu}q_{\mu}}p_{\mu}+\sqrt{1+\ell^2 p^{\mu}p_{\mu}} q_{\mu}-\ell \varepsilon_{\mu}^{\phantom{\mu}\nu\rho}p_{\nu}q_{\rho}~.
\end{equation}

The relevant DSR-deformed relativistic symmetries are particularly simple (with respect to other
much-studied examples~\cite{goldenrule}) since
the action of Lorentz-sector generators on momenta remains undeformed.
Indeed by posing
\begin{align}
\{R,p_0\}&=0 & \{N_1,p_0\}&=p_1 & \{N_2,p_0\}&=p_2\\
\{R,p_1\}&=-p_2 & \{N_1,p_1\}&=p_0 & \{N_2,p_1\}&=0\\
\{R,p_2\}&=p_1 & \{N_1,p_2\}&=0 & \{N_2,p_2\}&=p_0
\end{align}
one easily finds that the mass shell (\ref{mass-shellDOPPIONE})
is invariant and the composition law (\ref{composizione}) is covariant.

So we are dealing with a DSR-relativistic framework where the core aspect of the
deformation is the action of translation transformations on multiparticle states.
This was so far only left implicit by noticing that the momentum charges must be composed
following the nonlinear law (\ref{composizione}). But let us now notice that this implies
a deformed action of translations on multiparticles states. Take for example a system composed
of only two particles, respectively with phase-space coordinates $p_\mu,x^\nu$ and $q_\mu,y^\nu$:
then a translation parametrized by $b^\rho$,
and generated by the total-momentum charge $(p\oplus q)_\rho$,  acts for example
on the particle with phase-space coordinates $p_\mu,x^\nu$ as follows
\begin{equation}
b^\rho \{ (p\oplus q)_\rho ,x^{\nu}\}\simeq  b^\rho\{ p_\rho,x^{\nu}\}-\ell b^\rho\varepsilon_{\rho}^{\phantom{\rho}\sigma\gamma}q_\gamma\{ p_\sigma,x^{\nu}\}
\end{equation}
where on the right-hand side we were satisfied to show the leading-order Planck-scale modification.

Concerning translations acting on single-particle momenta
we notice that since our coordinates are such
that $ \{x^{\mu},x^{\nu}\}=\ell \varepsilon^{\mu\nu}_{\phantom{\mu\nu}\rho} x^{\rho}$
one could not possibly adopt the standard $\{p_{\mu},x^{\nu}\}=-\delta_\mu^\nu$
since then the Jacobi identities would not be satisfied
(in particular $\{p_{\mu},\{x^{\nu},x^{\rho}\}\}+
\{x^{\rho},\{p_{\mu},x^{\nu}\}\}+\{x^{\nu},\{x^{\rho},p_{\mu}\}\}\neq0$).
We adopt the description of
translations acting on single-particle momenta
given by
\begin{equation}
\label{poisson-deformato}
 \{p_{\mu},x^{\nu}\}=-\delta_{\mu}^{\nu}\sqrt{1+\frac{\ell^2}{4}p^{\rho}p_{\rho}}+\frac{\ell}{2} \varepsilon_{\mu}^{\phantom{\mu}\nu\rho}p_{\rho}
\end{equation}
which does satisfy the Jacobi identities.

One can easily verify that also (\ref{poisson-deformato})
and (\ref{pncc}) are compatible with undeformed rule of action of Lorentz transformations, {\it i.e.}
 the relevant Jacobi identities are satisfied assuming
\begin{align}
\{R,x^0\}&=0 & \{N_1,x^0\}&=-x^1 & \{N_2,x^0\}&=-x^2\\
\{R,x^1\}&=-x^2 & \{N_1,x^1\}&=-x^0 & \{N_2,x^1\}&=0\\
\{R,x^2\}&=x^1 & \{N_1,x^2\}&=0 & \{N_2,x^2\}&=-x^0~.
\end{align}

\section{Describing dynamics within the relative-locality framework}\label{secframework}
Up to this point we focused on a description of the momentum-space and spacetime structures of the theory here
of interest, with emphasis on relativistic implications. Our next task is to describe particle dynamics governed by these structures.
We shall be here satisfied with a description of dynamics in the classical limit,
but this is already a severe challenge,
particularly because of the implications of momentum-space curvature.
A formalism suitable for our purposes
was only recently developed: this is the relative-locality curved-momentum-space
framework of Refs.~\cite{prl,grf2nd}.
The main objective of this section is to formulate a
relative-locality curved-momentum-space theory which incorporates
the 3D-gravity structures we discussed.

\subsection{On-shell relation and relative-locality geometry of momentum space}
In Sec.~\ref{secgeometry} we established that the metric of the momentum
space of 3D gravity is the anti-de Sitter-space metric.
And we also commented briefly previous results indicating that
the momenta are governed by an on-shell condition
of the form
\begin{equation}
 \ell^{-2}\left(\arcsin\left(\sqrt{-\ell^2p^{\mu}p_{\mu}}\right)\right)^2=m^2~.
 \label{jocMMM}
\end{equation}

In the relative-locality framework, which we shall use for formulating dynamics,
the momentum-space metric and the on-shell relation
must be linked~\cite{prl,grf2nd} by the requirement that
the on-shell relation be describable in terms
the geodesic distance $D(0,{\bf p})$ of the momentum ${\bf p}$ from the origin:
\begin{equation}
\label{mass-shell}
 D^2(0,{\bf p})=m^2.
\end{equation}
In this subsection we verify that this requirement enforced by the relative-locality
framework reproduces the expected result (\ref{jocMMM}).
For this purpose we exploit the fact that our anti-deSitter
momentum space can be embedded very easily in $\mathbb{R}^{2,2}$,
\begin{equation}
\label{R2,2}
 ds^2=-du^2-(d\xi_0)^2+(d\xi_1)^2+(d\xi_2)^2~.
\end{equation}
By embedding our anti-de Sitter momentum space in $\mathbb{R}^{2,2}$
we can then describe the metric on the anti-de Sitter momentum space
as a metric induced by  the $\mathbb{R}^{2,2}$ metric.
 Our embedding coordinates are $Y_{I}=(\sqrt{1+\ell^2 p^{\mu}p_{\mu}},\ell p_{\mu})$ so
  we can evidently describe
the pull-back of the metric (\ref{R2,2}) to our $SL(2,R)$ as follows:
\begin{equation}
\label{metrica-spazio-momenti}
 ds^2=-(dp_0)^2+(dp_1)^2+(dp_2)^2-\frac{\ell^2 p^{\mu}p^{\nu}dp_{\mu}dp_{\nu}}{1+\ell^2 p^{\mu}p_{\mu}}~.
\end{equation}

In order to characterize
the geodesic distance $D(0,{\bf p})$
it is convenient  to adopt the description of the relevant geodesics
from the viewpoint of the embedding space $\mathbb{R}^{2,2}$.
We start
 by noticing that a geodesic on the anti-deSitter hypersurface (which is the image of our embedding), equipped with the Levi-Civita connection associated to the metric (\ref{metrica-spazio-momenti}),
can be described by the Lagrangian
\begin{equation}
\label{geodedica}
L=\dot{Y}^{I}\dot{Y}_{I}+\lambda(Y^{I}Y_{I}+1).
\end{equation}
The kinetic term
describes the free motion in $\mathbb{R}^{2,2}$ and $\lambda$ is a Lagrange multiplier
imposing that the motion should be on the anti-deSitter hypersurface.
The equations of motion one derives from the Lagrangian (\ref{geodedica}) are simply
\begin{equation}
 \begin{array}{l}
   \ddot{Y_I}=\lambda Y_I\\
   Y^IY_I+1=0.
 \end{array}
\end{equation}
The first equation is a simple second order differential equation, while the second one defines the anti-deSitter
hypersurface.

The geodesics going out from
the origin and arriving at a point $Y^I=(\sqrt{1+\ell^2 p^{\mu}p_{\mu}},\ell p_{\mu})$
are strongly characterized by the value of $p^{\mu}p_{\mu}$.
If $p^{\mu}p_{\mu}>0$ the geodesic is space-like and we have (taking the absolute value
of $\dot{Y}^{I}\dot{Y}_{I}$
when computing the geodesic distance)
\begin{equation}
 \ell^2 p^{\mu}p_{\mu}=\sinh^2(D(0,Y)).
\end{equation}
If  $p^{\mu}p_{\mu}=0$ the geodesics is light-like and $D(0,Y)=0$. Finally, if $p^{\mu}p_{\mu}<0$ the geodesic is
time-like and we have
\begin{equation}
 \ell^2 p^{\mu}p_{\mu}=-\sin^2(D(0,Y)).
\end{equation}
Using that $D(0,Y)=\ell D(0 ,{\bf p})=\ell m$ we can rewrite the previous equations as
\begin{align}
  \ell^2p^{\mu}p_{\mu}&=\sinh^2(\ell m)     &  p^{\mu}p_{\mu}&>0 \\
   p^{\mu}p_{\mu}&=0                        &  p^{\mu}p_{\mu}&=0 \\
   \ell^2 p^{\mu}p_{\mu}&=-\sin^2(\ell m)   &  p^{\mu}p_{\mu}&<0.
 \end{align}
Since our mass-shell condition should be a perturbation of the special relativistic one,
the physically relevant cases are the last two, that can be written together as
\begin{equation}
\label{geodistance}
 \ell ^2p^{\mu}p_{\mu}=-\sin^2(\ell m)
\end{equation}
where m is now allowed to be 0.

We notice that our physical momentum space is then described by the condition
\begin{equation}
-\ell^{-2}\leq p^{\mu}p_{\mu}\leq 0,
\end{equation} where the first inequality comes from the anti-deSitter nature
of our momentum space and the second one comes from the requirement for the mass-shell condition
to have the right special relativistic limit.

Rewriting (\ref{geodistance}) in the spirit of (\ref{mass-shell}), we have
\begin{equation}
 \ell^{-2}\left(\arcsin\left(\sqrt{-\ell^2p^{\mu}p_{\mu}}\right)\right)^2=m^2~,
 \label{jocnn}
\end{equation}
which, as announced, reproduces the prediction (\ref{jocMMM})
based on previous 3D-gravity results.

\subsection{Spinning affine connection on momentum space}
In the characterization of the geometry of momentum
space adopted in the relative-locality framework one combines
information on the metric of momentum space (in the sense discussed in the
previous subsection) with a specification of
the connection coefficients on momentum space, which must be based~\cite{prl,grf2nd}
on the form of the law of composition
of momenta near the origin of momentum space.
Near the
origin of momentum space
the composition law relevant for our spinning-spacetime case
takes the form
\begin{equation}
\label{composizioneLEAD}
(p \oplus q)_{\mu} \simeq p_{\mu}+ q_{\mu}-\ell \varepsilon_{\mu}^{\phantom{\mu}\nu\rho}p_{\nu}q_{\rho}+\frac{\ell^2}{2}(q^\nu q_\nu p_\mu+ p^\nu p_\nu q_\mu)~.
\end{equation}
Following Refs.~\cite{prl,grf2nd} one must determine the connection coefficients $ \Gamma^{\nu\rho}_{\mu}(0)$
from the leading-order term of the expansion of the composition law near the
origin of momentum space:
\begin{equation}
\label{conncoeff}
(p \oplus q)_{\mu} \simeq p_{\mu}+ q_{\mu}
- \ell \Gamma_{\mu}^{\phantom{\mu}\nu\rho}(0)p_{\nu}q_{\rho} + \dots
\end{equation}
(adopting conventions~\cite{grf2nd} such that the connection coefficients are
dimensionless).

Therefore in our case the connection coefficients are
\begin{equation}
 \Gamma^{\nu\rho}_{\mu}(0)=\varepsilon_{\mu}^{\phantom{\mu}\nu\rho}~.
\end{equation}

Also part of the notion of  ``relative-locality momentum-space geometry"
introduced in Refs.~\cite{prl,grf2nd}
are definitions for torsion, curvature of the connection and nonmetricity.
Let us see these directly in action for the momentum space here of interest.
Our momentum space has torsion
\begin{equation}
T_{\mu}^{\nu \rho}(0)=-\frac{\partial }{\partial p_{\nu}}\frac{\partial }{\partial q_{\beta}}((p\oplus q)_\mu-(q \oplus p)_{\mu})_{p=q=0}=2 \ell \Gamma_{\mu}^{[\nu \rho]}(0)=\ell \Gamma_{\mu}^{\nu \rho}(0)-\Gamma_{\mu}^{ \rho\nu}(0)=2\ell \varepsilon_{\mu}^{\phantom{\mu}\nu\rho}~,
\end{equation}
but the curvature of the connection  (evaluated in the origin) vanishes
\begin{equation}
R^{\nu \rho \sigma}_{\mu}(0)=2\frac{\partial }{\partial p_{[\nu}}\frac{\partial }{\partial q_{\rho]}}\frac{\partial }{\partial k_{\sigma}} ((p \oplus q) \oplus k-p \oplus (q \oplus k))_{\mu}\mid_{p=q=k=0}=0.
\end{equation}
Finally we determine the value in the origin of the nonmetricity tensor~\cite{prl,grf2nd} in
terms of the spinning affine connection here discussed and the metric
of anti-de Sitter momentum space
discussed in Sec.~\ref{secgeometry}, finding that it vanishes:
\begin{equation}
N^{\rho \mu \nu}(0)=\nabla^{\rho} g^{\mu \nu}(0)= g^{\mu \nu,\rho}(0)+\ell \Gamma_{\sigma}^{\mu \rho}(0)g^{\sigma \nu}(0)+\ell \Gamma_{\sigma}^{\nu \rho}(0)g^{\mu \sigma}(0)
= 0.
\nonumber
\end{equation}

\subsection{Classical regime within the relative-locality framework}
As announced we shall focus on dynamics in the classical limit,
as formalized within the relative-locality
framework of Refs.~\cite{prl,grf2nd}. Accordingly we shall formulate interactions
among particles through boundary terms at endpoints of wordlines
enforcing momentum conservation. And as previous relative-locality-framework
studies we shall be satisfied with results obtained at leading order in $\ell$.
For example the case of a single two-body-particle-decay process
(see Fig.~1)
is then described in terms of the following action~\cite{prl,grf2nd,anatomy}:
\begin{equation}
\begin{array}{lll}
\mathcal{S}	&=& \displaystyle\int_{-\infty}^{s_0}((\delta^{\mu}_{\nu}-\frac{\ell}{2}\epsilon^{\mu\sigma}_{\phantom{\mu\sigma}\nu}k_{\sigma})z^\nu\dot{k}_{\mu}+\mathcal{N}_k(k^{\mu}k_{\mu}-m^2))ds+
\int_{s_{0}}^{\infty}((\delta^{\mu}_{\nu}-\frac{\ell}{2}\epsilon^{\mu\sigma}_{\phantom{\mu\sigma}\nu}p_{\sigma})x^\nu\dot{p}_{\mu}+\mathcal{N}_p (p^{\mu}p_{\mu}-m^{\prime \, 2}))
ds\\
			& &\displaystyle+\int_{s_{0}}^{\infty}((\delta^{\mu}_{\nu}-\frac{\ell}{2}\epsilon^{\mu\sigma}_{\phantom{\mu\sigma}\nu}q_{\sigma})y^\nu\dot{q}_{\mu}+\mathcal{N}_q (q^{\mu}q_{\mu}-m^{\prime \, 2}))ds -\xi_{[0]}^{\mu}\mathcal{K}_{\mu}^{[0]}(s_{0}) \ .
\end{array}
\label{joctra}
\end{equation}
Here the Lagrange multipliers $\mathcal{N}_k,\mathcal{N}_p ,\mathcal{N}_q$
enforce in standard way the on-shellness of particles, denoting with $m$ the mass of the incoming particle
 and denoting with $m^\prime$ and $m^{\prime\prime}$ the masses of the
 outgoing particles.
Also notice that the simplectic form was specialized to the case of spacetime coordinates
such that $\{x_{\mu},x_{\nu}\}=\ell \varepsilon_{\mu\nu}^{\phantom{\mu\nu}\rho} x_{\rho}$,
so that then (see (\ref{poisson-deformato})) one must be enforce
\begin{equation}
\label{poisson-deformatoSHbis}
 \{p_{\mu},x^{\nu}\}=-\delta^{\nu}_{\mu}+\frac{\ell}{2} \varepsilon_{\mu}^{\phantom{\mu}\nu\sigma}p_{\sigma}~.
\end{equation}
And the most innovative part of the formalization introduced in Refs.~\cite{prl,grf2nd}
is the presence of boundary terms at endpoints of wordlines
enforcing momentum conservation, such
as the one characterized in (\ref{joctra}) by $\mathcal{K}_{\mu}^{[0]}(s_{0})$.
In particular, on the basis of what we established in the previous section
we can specify the boundary term in (\ref{joctra}) as follows
\begin{equation}
\begin{array}{c}
\mathcal K_{\mu}^{[0]}(s_0)=(k)_\mu-(p\oplus q)_\mu =
k_\mu-p_{\mu}-q_{\mu}+\ell \varepsilon_{\mu}^{\phantom{\mu}\nu\rho}p_{\nu}q_{\rho}.
\end{array}
\label{consgac}
\end{equation}

\begin{figure}[htbp]
\begin{center}
\includegraphics[scale=0.6]{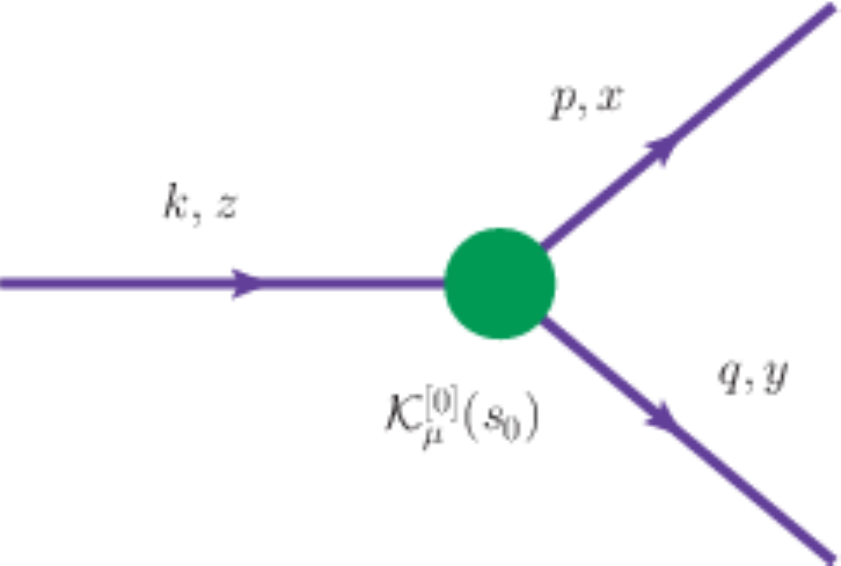}
\end{center}
\caption{We here show the interaction described in Eq.~(\ref{joctra}), with one incoming and
two outgoing particles.}
\label{joc1}
\end{figure}

Many significant implications of the momentum-space curvature are a consequence of the way in which
translational invariance manifests itself, as stressed already in Sec.~\ref{secdsr}.
Translations are still generated by the total-momentum charges, but these are obtained from
single-particle charges via the deformed $\oplus$ composition law.
And we shall find that, as in similar analyses
of the relative-locality framework~\cite{prl,grf2nd,anatomy,leelaurentGRB},
this produces relativity of spacetime locality.
Actually relativity of spacetime locality appears to be
a generic consequence of a nontrivial geometry
of momentum space, which already affects such theories for
free particles~\cite{whataboutbob}, when distantly boosted observers are considered.
It amounts to the possibility that pairs of events
established to be coincident
by nearby observers may be described as events that are not exactly coincident in the coordinatizations
of those events by
distant observers~\cite{prl,grf2nd,whataboutbob}.
Because of these coordinate artifacts one cannot trust the description of a given observer Alice
of events distant from her: one must in such cases replace Alice description with the coordinatization
of the events by some observer Bob near to them. The same physical content one usually
({\it i.e.} with trivial translation transformations) produces
by simply deriving the equations of motion, here requires us to handle both the equations
of motion and the laws of transformation among distant observers. We shall see an example
of this mechanism explicitly in the next section.

\section{Dual-gravity lensing}
Our last task is to expose one aspect of the relativity of spacetime locality
present in out 3D-gravity-inspired theory.
For this it will suffice to consider an example of causally-connected
interactions, such as the one
 here shown in Fig.~2.

\begin{figure}[htbp]
\begin{center}
\includegraphics[scale=0.55]{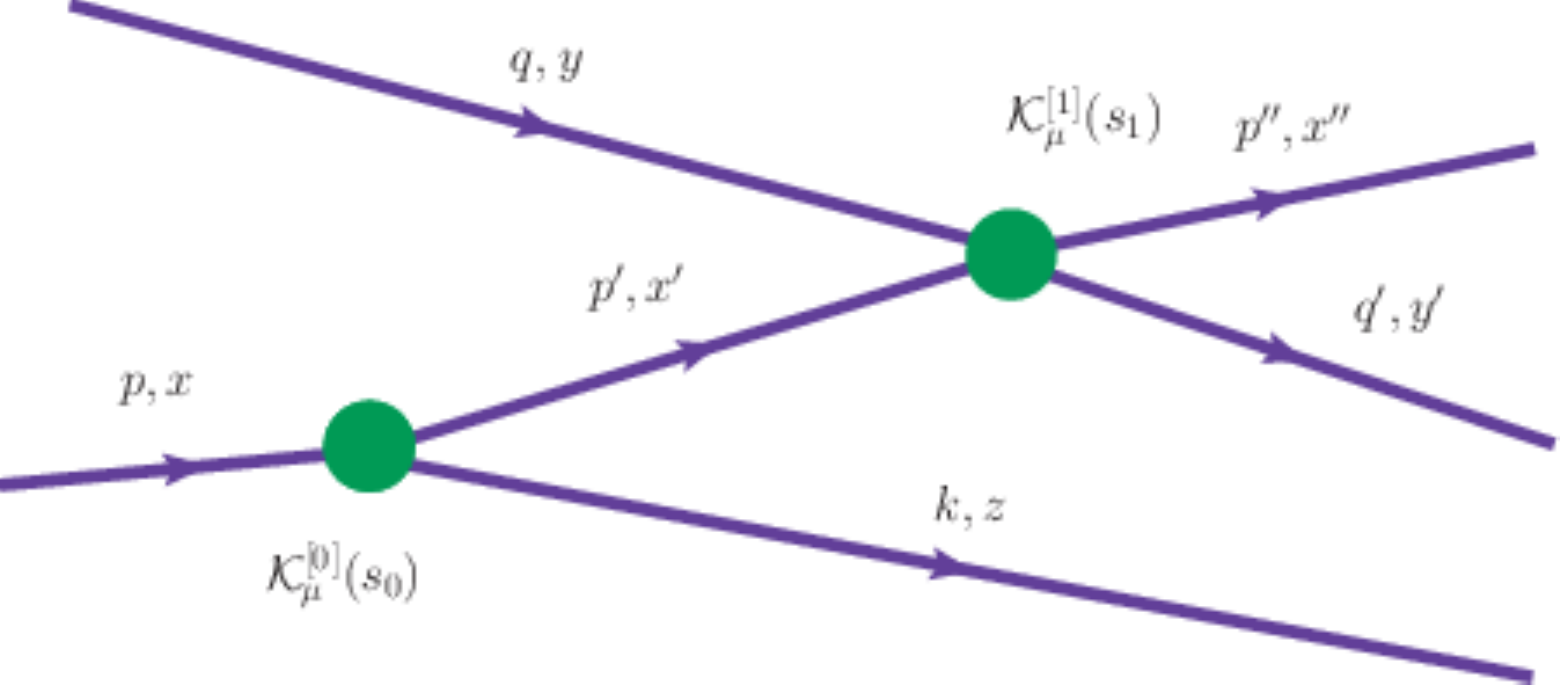}
\end{center}
\caption{We here show the process described in Eq.~(\ref{action}), involving two causally-connected
interactions.}
\label{joc2}
\end{figure}

The situation in Fig.~2 is described within the relative-locality framework by an action of the type
\begin{equation}\label{action}
\begin{array}{lll}
\mathcal S	&=&\displaystyle \int_{-\infty}^{s_0}((\delta^{\mu}_{\nu}-\frac{\ell}{2}\epsilon^{\mu\sigma}_{\phantom{\mu\sigma}\nu}p_{\sigma})x^\nu \dot{p}_\mu +\mathcal N_p (p^\mu p_\mu-m^{2}))ds
+\int_{s_0}^{+\infty}((\delta^{\mu}_{\nu}-\frac{\ell}{2}\epsilon^{\mu\sigma}_{\phantom{\mu\sigma}\nu}k_{\sigma})z^\nu \dot{k}_\mu +\mathcal N_k (k^\mu k_\mu-m^{\prime \, 2}))ds+\\
						& &\displaystyle +\int_{s_0}^{s_1}((\delta^{\mu}_{\nu}-\frac{\ell}{2}\epsilon^{\mu\sigma}_{\phantom{\mu\sigma}\nu}p'_{\sigma})x'^\nu \dot{p'}_\mu
+\mathcal N_{p'}(p^{\prime \, \mu}p^{\prime}_{\mu}-m^{\prime\prime \, 2}))ds+\int_{-\infty}^{s_1}((\delta^{\mu}_{\nu}-\frac{\ell}{2}\epsilon^{\mu\sigma}_{\phantom{\mu\sigma}\nu}q_{\sigma})y^\nu \dot{q}_\mu +\mathcal N_q (q^{\mu}q_\mu-\mu^{2}))ds+\\
						& &\displaystyle +\int_{s_1}^{+\infty}((\delta^{\mu}_{\nu} -\frac{\ell}{2}\epsilon^{\mu\sigma}_{\phantom{\mu\sigma}\nu}p''_{\sigma})x''^\nu \dot{p''}_\mu +\mathcal N_{p''}(p^{\prime\prime \, \mu}p^{\prime\prime}_{\mu})-\mu^{\prime \, 2}))ds
+\int_{s_1}^{+\infty}((\delta^{\mu}_{\nu} -\frac{\ell}{2}\epsilon^{\mu\sigma}_{\phantom{\mu\sigma}\nu}q'_{\sigma})y'^\nu \dot{q'}_\mu +\mathcal N_{q'}(q^{\prime \, \mu}q^{\prime}_{\mu}-\mu^{\prime\prime \, 2}))ds+\\
						& &\displaystyle -\xi_{(0)}^\mu \mathcal K_{\mu}^{[0]}(s_0)-\xi_{(1)}^\mu \mathcal K_{\mu}^{[1]}(s_1)
\end{array}
\end{equation}
where we gave each particle possibly a different mass
($m$,$m^{\prime}$,$m^{\prime\prime}$,$\mu$,$\mu^{\prime}$,$\mu^{\prime\prime}$).
Concerning the conservation laws,
which are to be codified in
the boundary terms $\mathcal K_{\mu}^{[0]}(s_0)$, $\mathcal K_{\mu}^{[1]}(s_1)$,
we follow the prescription given in Ref.~\cite{anatomy} for having causally-connected
interactions preserving translational invariance, so we adopt as boundary terms
\begin{equation}\label{1conslaw}
\begin{array}{c}
\mathcal K_{\mu}^{[0]}(s_0)=(q\oplus p)_\mu-(q\oplus p'\oplus k)_\mu =p_\mu-p'_\mu-k_\mu-\ell\varepsilon_\mu^{\alpha\beta}(q_\alpha p_\beta-q_\alpha p'_\beta-q_\alpha k_\beta-p'_\alpha k_\beta)
\end{array}
\end{equation}
and
\begin{equation}\label{2conslaw}
\begin{array}{c}
\displaystyle \mathcal K_{\mu}^{[1]}(s_1)=(q\oplus p'\oplus k)_\mu-(p''\oplus q'\oplus k)_\mu
 =q_\mu+p'_\mu-p''_\mu-q'_\mu-\ell\varepsilon_\mu^{\alpha\beta}(q_\alpha p'_\beta+q_\alpha k_\beta+p'_\alpha k_\beta-p''_\alpha q'_\beta-p''_\alpha k_\beta-q'_\alpha k_\beta).
\end{array}
\end{equation}

We are now all set to derive
equations of motion and boundary conditions,
by varying  (\ref{action}) keeping momenta
fixed~\cite{prl,grf2nd} at $\pm\infty$.
For the equations of motion one easily finds
\begin{equation}
\dot{p}_\mu=0,\hspace{5pt}\dot{q}_\mu=0,\hspace{5pt}\dot{q'}_\mu=0,\hspace{5pt}\dot{k}_\mu=0,\hspace{5pt}\dot{p'}_\mu=0,\hspace{5pt}\dot{p''}_\mu=0,
\label{jocEOM1}
\end{equation}

\begin{equation}
\mathcal C_p=0,\hspace{5pt} \mathcal C_q=0,\hspace{5pt} \mathcal C_{q'}=0,\hspace{5pt} \mathcal C_k=0,\hspace{5pt} \mathcal C_{p'}=0,\hspace{5pt} \mathcal C_{p''}=0,\hspace{5pt}
\label{jocEOM2}
\end{equation}

\begin{equation}
\dot{x}^\mu=2\mathcal N_p p^\mu,\hspace{5pt} \dot{y}^\mu=2\mathcal N_qq^\mu,\hspace{5pt} \dot{y'}^\mu=2\mathcal N_{q'} q^{\prime \, \mu},
\label{jocEOM3}
\end{equation}

\begin{equation}
\dot{z}^\mu=2\mathcal N_kk^{\mu},\hspace{5pt} \dot{x'}^\mu=2\mathcal N_{p'}p^{\prime \, \mu},\hspace{5pt} \dot{x''}^\mu=2\mathcal N_{p''}p^{\prime\prime \, \mu},
\label{jocEOM4}
\end{equation}
and the boundary conditions at endpoints of worldlines are

\begin{equation}
\begin{array}{l}
\displaystyle z^\mu(s_0)=-\xi^\nu_{[0]}\frac{\delta \mathcal K_\nu^{[0]}}{\delta k_\sigma}(\delta^{\mu}_{\sigma}+\frac{\ell}{2}\varepsilon^{\mu\phantom{\sigma}\rho}_{\phantom{\mu}\sigma}k_{\rho})=
\xi_{[0]}^\mu-\frac{\ell}{2}\varepsilon_\nu^{\phantom{\nu}\mu\alpha}(k_\alpha-2q_\alpha-2p^\prime_\alpha)\xi_{[0]}^\nu,\\

\displaystyle x^\mu(s_0)=\xi^\nu_{(0)}\frac{\delta \mathcal K_\nu^{(0)}}{\delta p_\sigma}(\delta^{\mu}_{\sigma}+\frac{\ell}{2}\varepsilon^{\mu\phantom{\sigma}\rho}_{\phantom{\mu}\sigma}p_{\rho})=
\xi_{[0]}^\mu-\frac{\ell}{2}\varepsilon_\nu^{\phantom{\nu}\mu\alpha}(p_\alpha-2q_\alpha)\xi_{[0]}^\nu,\\

\displaystyle x'^\mu(s_0)=-\xi^\nu_{[0]}\frac{\delta \mathcal K_\nu^{[0]}}{\delta p'_\sigma}(\delta^{\mu}_{\sigma}+\frac{\ell}{2}\varepsilon^{\mu\phantom{\sigma}\rho}_{\phantom{\mu}\sigma}p'_{\rho})=
\xi_{[0]}^\mu-\frac{\ell}{2}\varepsilon_\nu^{\phantom{\nu}\mu\alpha}(p^\prime_\alpha-2k_\alpha+2q_\alpha)\xi_{[0]}^\nu,\\

\displaystyle x'^\mu(s_1)=\xi^\nu_{[1]}\frac{\delta \mathcal K_\nu^{[1]}}{\delta p'_\sigma}(\delta^{\mu}_{\sigma}+\frac{\ell}{2}\varepsilon^{\mu\phantom{\sigma}\rho}_{\phantom{\mu}\sigma}p'_{\rho})=
\xi_{[1]}^\mu-\frac{\ell}{2}\varepsilon_\nu^{\phantom{\nu}\mu\alpha}(p^\prime_\alpha-2k_\alpha+2q_\alpha)\xi_{[1]}^\nu,\\

\displaystyle x''^\mu(s_1)=-\xi^\nu_{[1]}\frac{\delta \mathcal K_\nu^{[1]}}{\delta p''_\sigma}(\delta^{\mu}_{\sigma}+\frac{\ell}{2}\varepsilon^{\mu\phantom{\sigma}\rho}_{\phantom{\mu}\sigma}p''_{\rho})=
\xi_{[1]}^\mu-\frac{\ell}{2}\varepsilon_\nu^{\phantom{\nu}\mu\alpha}(p^{\prime\prime}_\alpha-2q^\prime_\alpha-2k_\alpha)\xi_{[1]}^\nu,\\

\displaystyle y^\mu(s_1)=\xi^\nu_{[1]}\frac{\delta \mathcal K_\nu^{[1]}}{\delta q_\mu}(\delta^{\mu}_{\sigma}+\frac{\ell}{2}\varepsilon^{\mu\phantom{\sigma}\rho}_{\phantom{\mu}\sigma}q_{\rho})=
\xi_{[1]}^\mu-\frac{\ell}{2}\varepsilon_\nu^{\phantom{\nu}\mu\alpha}(q_\alpha-2p^\prime_\alpha-2k_\alpha)\xi_{[1]}^\nu,\\

\displaystyle y'^\mu(s_1)=-\xi^\nu_{[1]}\frac{\delta \mathcal K_\nu^{[1]}}{\delta q'_\sigma}(\delta^{\mu}_{\sigma}+\frac{\ell}{2}\varepsilon^{\mu\phantom{\sigma}\rho}_{\phantom{\mu}\sigma}q'_{\rho})=
\xi_{[1]}^\mu-\frac{\ell}{2}\varepsilon_\nu^{\phantom{\nu}\mu\alpha}(q^\prime_\alpha-2k_\alpha+2p^{\prime\prime}_\alpha)\xi_{[0]}^\nu,\\

\displaystyle \mathcal K_\mu^{[0]}(s_0)=0,\hspace{5pt}\mathcal K_\mu^{[1]}(s_1)=0,
\frac{\delta \mathcal K_\mu^{[0]}}{\delta q_\nu}=0,\hspace{5pt}
\frac{\delta \mathcal K_\mu^{[1]}}{\delta k_\nu}=0.
\end{array}
\label{jocBOUNDARY}
\end{equation}

Evidently (and unsurprisingly) when only ``soft interactions"~\cite{anatomy}
are involved, {\it i.e.} all particles involved have energies small enough that
the $\ell$-deformation can be ignored, a standard special relativistic situation
is recovered. We are going to focus in particular on what the theory predicts for
the particle exchanged between the two interactions, assuming it is a
massless particle.
And we shall make reference to an observer Alice located where the
interaction with conservation law $\mathcal K_{\mu}^{[0]}=0$
takes place, and an observer Bob
located where the
interaction with conservation law $\mathcal K_{\mu}^{[1]}=0$ takes place.

If both interactions in Fig.~2 are soft then Alice describes the exchanged massless
particle according to
\begin{equation}
x^1_{A(s)}=x^0_{A(s)}~,~~~x^2_{A(s)}=0
\label{jocalice}
\end{equation}
where we assumed that the first interaction occurs exactly in Alice's origin,
and we further specialized conventions so that the massless particles exchanged
between two soft interactions propagates along a common $x^1$ axis of Alice and Bob.
The index $(s)$ introduced in (\ref{jocalice}) will be here consistently used to
identify equations written for a soft particle (a particle only taking part in interactions
for which the $\ell$-deformation is negligible).

Bob is distant and at rest with respect to Alice, and we want that the massless
particle exchanged between two soft interactions is detected, through the second
interaction, in Bob's origin. So Bob must be connected to Alice by a translation
of parameters $b^{\mu}=(b^0,b^1,0)$ with $b^0=b^1$.
Evidently the worldline described as in (\ref{jocalice}) according
to Alice's coordinatization
maintains that description in Bob's coordinatization:
\begin{equation}
x^1_B=x^0_B+b^{0}-b^1=x^0_B
~,~~~x^2_B=0
\label{jocbob}
\end{equation}

Of course, what we are interested in understanding is how this
situation changes if the processes are ``hard" enough for the $\ell$-deformation
to be tangible.
As those familiar with relative locality will be expecting (and newcomers will see here below),
for such ``hard" particles it is not even obvious at the onset of the analysis
which of them will reach Bob in his origin.
We shall find that some hard particles emitted in Alice origin do reach Bob's origin, but
this will come about only upon allowing ourselves to consider
hard particles not necessarily emitted along Alice's  $x^1$ axis.
We chose above notation such that the particle exchanged between the
interactions has phase-space coordinates  $p^{\prime}_{\mu}$,$x^{\prime \,\mu}$
and we shall keep using consistently this notation.
And we characterize the direction of propagation
of this exchanged particle according to Alice via
an angle $\theta$:
\begin{equation}
\begin{array}{l}
 p^{\prime \, 1}=p^{\prime \, 0}\cos\theta \\
 p^{\prime \, 2}=-p^{\prime \, 0}\sin\theta
\end{array}
\end{equation}
The case of one such particle emitted from Alice's origin would then be described
according to
\begin{equation}
\begin{array}{l}
\label{moto alice}
 x^{\prime \, 1}_A=x^{\prime \, 0}_A\cos\theta \\
 x^{\prime \, 2}_A=-x^{\prime \, 0}_A\sin\theta
\end{array}
\end{equation}
In order to establish if
any of these particles ({\it i.e.} if for any value of energy $p^{\prime \, 0}$
and emission angle $\theta$ with respect to Alice's   $x^1$ axis)
manages to reach Bob's origin we evidently need to determine Bob's description of
the worldlines described by Alice according to (\ref{moto alice}).
The crucial step for this is for us to establish the relationship
between Alice's and Bob's description of the coordinates of the particle exchanged between
the two interactions in Fig.~2, {\it i.e.} we need to perform a translation transformation
properly taking into account the deformed law of composition of momentum charges.

Following previous results on translation symmetry in the relative-locality
framework~\cite{prl,grf2nd,anatomy} (see also Ref.~\cite{cortesRL})
we implement the relevant translation transformations through the action of the total-momentum charge:
\begin{eqnarray}
 x^{\prime \, \mu}_B &=& x^{\prime \, \mu}_A+b^{\nu}\{(q\oplus p'\oplus k)_{\nu},x^{\prime \, \mu}\}=x^{\prime \, \mu}_A+b^{\nu}\frac{\partial (q\oplus p'\oplus k)_{\nu}}{\partial p'_{\sigma}}(\delta^{\mu}_{\sigma}
+\frac{\ell}{2}\varepsilon^{\mu\phantom{\sigma}\rho}_{\phantom{\mu}\sigma}p'_{\rho})=\nonumber\\
&=& x^{\prime \, \mu}_A+b^{\nu}(-\delta_{\nu}^{\sigma}+\ell\varepsilon_{\nu}^{\phantom{\nu}\sigma\gamma}(k_\gamma-q_\gamma))(\delta^{\mu}_{\sigma}
+\frac{\ell}{2}\varepsilon^{\mu\phantom{\sigma}\rho}_{\phantom{\mu}\sigma}p'_{\rho})=x^{\prime \, \mu}_A -b^{\mu}+b^{\nu}\frac{\ell}{2}\varepsilon_{\nu}^{\phantom{\nu}\mu\rho}(p'_{\rho}+2k_\rho-2q_\rho)
\equiv x^{\prime \, \mu}_A- \Delta^{\mu}
\label{traslxprime}
\end{eqnarray}
Analogously we find the translations for the other coordinates:
\begin{equation}
\begin{array}{l}
\displaystyle x^\mu_B=x^\mu_A+b^\nu\{(q\oplus p)_\nu,x^\mu_A\}=x^\mu_A-b^\mu+\frac{\ell}{2}\varepsilon_\nu^{\phantom{\nu}\mu\rho}(p_\rho-2q_\rho)b^\nu\\

\displaystyle y^\mu_B=y^\mu+b^\nu\{(q\oplus p'\oplus k)_\nu,y^\mu_A\}=y^\mu_A-b^\mu+\frac{\ell}{2}\varepsilon_\nu^{\phantom{\nu}\mu\rho}(q_\rho+2p'_\rho+2k_\rho)b^\nu\\

\displaystyle z^\mu_B=z^\mu_A+b^\nu\{(q\oplus p'\oplus k)_\nu,z^\mu_A\}=z^\mu_A-b^\mu+\frac{\ell}{2}\varepsilon_\nu^{\phantom{\nu}\mu\rho}(k_\rho-2q_\rho-2p'_\rho)b^\nu\\

\displaystyle x''^\mu_B=x''^\mu_A+b^\nu\{(p''\oplus q'\oplus k)_\nu,x''^\mu_A\}=x''^\mu_A-b^\mu+\frac{\ell}{2}\varepsilon_\nu^{\phantom{\nu}\mu\rho}(p''_\rho+2q'_\rho+2k_\rho)b^\nu\\

\displaystyle y'^\mu_B=y'^\mu_A+b^\nu\{(p''\oplus q'\oplus k)_\nu,y'^\mu_A\}=y'^\mu_A-b^\mu+\frac{\ell}{2}\varepsilon_\nu^{\phantom{\nu}\mu\rho}(q'_\rho+2k_\rho-2p''_\rho)b^\nu
\end{array}
\end{equation}
Consistently with what had been previously established~\cite{prl,grf2nd,anatomy} in the literature on
translation symmetry in the relative-locality framework, one can easily verify that these
translation transformations produced by the total-momentum charge leave the equations of motion
(\ref{jocEOM1}), (\ref{jocEOM2}), (\ref{jocEOM3}), (\ref{jocEOM4}) and the boundary conditions (\ref{jocBOUNDARY}) unchanged.
As announced above our main focus is going to be on the case of the particle
with coordinates $x^{\prime \, \mu}$ which is exchanged between the two interactions
in Fig.~2.
On the basis of (\ref{traslxprime}) one finds that for the worldline
of this particle,
described by Alice according to (\ref{moto alice}), Bob's description is
\begin{eqnarray}
x_B^{\prime \, 1}=\cos \theta\ x_B^{\prime \, 0}+\cos \theta \Delta^0-\Delta^1 \label{bobhardone}\\
x_B^{\prime \, 2}=-\sin \theta x_B^{\prime \, 0}-\sin \theta \Delta^0-\Delta^2 ~, \label{bobhardtwo}
\end{eqnarray}
where $\Delta^\mu$ was introduced in  (\ref{traslxprime}): $\displaystyle \Delta^\mu \equiv b^{\mu}-\frac{\ell}{2}\varepsilon_{\nu}^{\phantom{\nu}\mu\rho}(p'_\rho+2k_\rho-2q_\rho)b^{\nu}$.

We cannot {\it a priori} insist on having
the hard photon reach Bob in his spacetime origin, but we can
enforce (at least for some choices of $\theta$
and of the translation parameters $b^\mu$) that the hard photon goes through
Bob's spatial origin $x_B^{\prime \, 1}=x_B^{\prime \, 2}=0$.
We observe that the equation of motion
can be easily rearranged as follows
\begin{equation}
\begin{array}{l}
\displaystyle x_B^{\prime \, 2}=-\tan\theta x_B^{\prime \, 1}-\tan\theta \Delta^1-\Delta^2\\
\displaystyle x_B^{\prime \, 0}=\frac{x_B^{\prime \, 1}-\cos \theta \Delta^0+\Delta^1}{\cos \theta}
\end{array}
\end{equation}
Enforcing that the particle goes through $x_B^{\prime \, 1}=x_B^{\prime \, 2}=0$
we have that
\begin{equation}
\begin{array}{l}
\displaystyle \tan\theta =-\frac{\Delta^2}{\Delta^1}\\
\displaystyle x_B^{\prime \, 0}=\frac{\Delta^1}{\cos \theta}-\Delta^0
\end{array}
\label{twoeqs}
\end{equation}
The first of these two equations leads to conclude that
\begin{equation}
\theta \simeq \tan\theta =-\frac{\Delta^2}{\Delta^1}=-\frac{b\ell(k_1-q_1+k_0-q_0)+b\frac{\ell}{2}(p'_0+p'_1)}{-b-b\ell(k_2-q_2)-b\frac{\ell}{2}p'_{2}}\approx \frac{b\ell(k_1-q_1+k_0-q_0)}{b+b\ell(k_2-q_2)} \simeq \ell (k_1-q_1+k_0-q_0)~,
\label{dualgravmagnitude}
\end{equation}
where we used\footnote{Note in particular that a term going like $\ell p'_{2}$ should only be be
taken into account at next-to-leading order, since from (\ref{bobhardtwo}) one
sees that $\ell p'_{2} \simeq \ell \theta p'_{0}$ and $\theta$
vanishes at zero-th order in $\ell$.
Similarly from  (\ref{bobhardone}) one sees
that $p'_0+p'_1 \simeq (1-\cos \theta) p'_0 \simeq \theta^2 p'_0$.}
the fact that $\theta$ vanishes at zero-th order in $\ell$  and we are working
at leading order in $\ell$.

Our result (\ref{dualgravmagnitude}) is evidently noteworthy: we found
a non-zero result for $\theta$, {\it i.e.}  the
worldlines of hard
particles that reach Bob from Alice must not be parallel to the worldlines
of the soft particles that reach Bob from Alice.
This is the feature known as ``dual-gravity lensing" in the relative-locality
literature~\cite{leelaurentGRB,transverse}.
Let us postpone further comments on this until after we have established
the time at which such hard particles cross Bob's spatial origin.
This is easily done by substituting our result for $\theta$
in the second of Eqs.~(\ref{twoeqs}):
\begin{equation}
x_B^{\prime \, 0}=\frac{\Delta^1}{\cos \theta}-\Delta^0\approx \Delta^1-\Delta^0\approx b+ b(k_2-q_2)-b-b(k_2-q_2)=0~.
\end{equation}
So we have that the relevant hard particles do cross the origin of Bob's reference
frame. In setting up the analysis we only had enough handles to specialize
to the case of hard particles going through Bob's spatial origin, but then the
result is such that those actually go through Bob's spacetime origin.

 Fig.~3 summarizes the findings of our analysis of massless particles exchanged
 between distant observers Alice and Bob.

\begin{figure}[htbp]
\begin{center}
\includegraphics[scale=0.6]{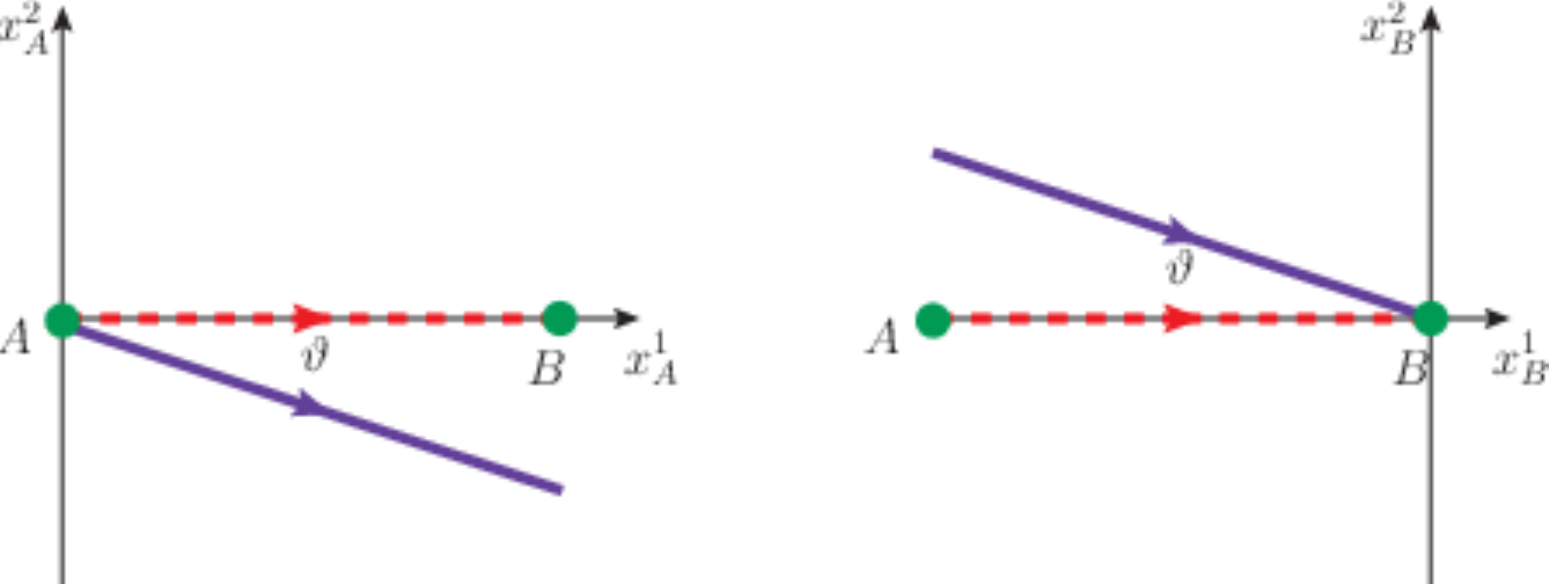}
\end{center}
\caption{We here summarize schematically the findings of our analysis of massless particles exchanged
 between distant observers Alice and Bob. The $x$ axis in figure is determined by the direction
 of the soft (low-energy) massless particle emitted at Alice that reaches Bob.
 Hard (high-energy) particles emitted at Alice  that also reach Bob are the ones that Alice describes
 as going along a direction forming an angle $\theta$ (whose value is determined by
 our Eq.~(\ref{dualgravmagnitude}) with the $x$ axis.
 We drew a macroscopic angle $\theta$ for better visibility, but actually this angle is extremely small
 (even if the particle energies involved are as high as, say, $1 TeV$ the angle $\theta$ still
only is of order $10^{-16}$).}
\label{joc2}
\end{figure}

Evidently the aspects of  ``dual-gravity lensing" shown in Fig.~3 can deserve a few
extra comments.
Like previous cases in which ``dual-gravity lensing" was encountered~\cite{leelaurentGRB,transverse},
we observe that relative locality plays a key role.
Let us focus for example on the event where the soft (red) worldline crosses
Bob's worldline
and the event where the hard (blue) worldline crosses the
line $x_B^{\prime \, 2}= x_B^{\prime \, 0}=0$.
These two events are coincident, as manifest in the coordinatization by the nearby observer Bob,
but Alice's  inferences about these two events, which are distant from Alice,
would describe them as noncoincident.

In previous related studies~\cite{leelaurentGRB,transverse}
one also finds an implicit invitation to study the energy dependence of dual-gravity lensing.
In our case the angle $\theta$ that governs the magnitude of the lensing is of order $\ell E_\star$,
where $E_\star$ is a characteristic energy scale of the process involved.
And it is interesting to compare what is expected as difference between a case with some $E_\star$
and a case with some $E'_\star$ bigger than $E_\star$.
The study of dual-gravity lensing reported in Ref.~\cite{leelaurentGRB} found that essentially
the relative angle of lensing, $\theta' - \theta$, would have to be proportional
to the sum of the energy scales involved, $E'_\star + E_\star$.
Instead the study reported in Ref.~\cite{transverse}
would predict in such cases a relative lensing effect going like
difference of the energy scales involved, $E'_\star - E_\star$.
Evidently the case we here uncovered is in the same class as the one of Ref.~\cite{transverse},
since indeed in our case the angle $\theta$ goes like  $\ell E_\star$,
and therefore $\theta' - \theta \approx \ell E_\star^{\prime} - \ell E_\star$.

\section{Outlook}~\label{closingsec}
We feel that the study we here reported should be viewed as confirming
the usefulness of techniques developed by research on DSR-deformed relativistic symmetries
for the analysis of scenarios motivated by the 3D-quantum-gravity literature.

And surely reference to results derived within an actual quantum-gravity model
(in spite of being only a 3D model) gives poignancy to DSR studies.
In this respect our result for dual-gravity lensing in the 3D-gravity-inspired scenario
has significance from a broader DSR perspective. Previous studies of 
dual-gravity lensing~\cite{leelaurentGRB,transverse} had provided possible alternative
pictures of dual-gravity lensing, and it is valuable to see which one results from
an actual (3D) quantum-gravity theory.

The most significant limitation of our analysis comes from neglecting quantum aspects
of the DSR and relative-locality features. But the first example of analysis
of such quantum effects is only very recent, reported in Ref.~\cite{fuzzy1},
and relies heavily on the fact basis
developed in nearly 20 years of investigations
of $\kappa$-Minkowski
noncommutativity. The much younger literature on the spinning spacetime here of interest
 still does not offer some of the ingredients
used in Ref.~\cite{fuzzy1} for the description of quantum effects. However, we feel that particularly
the spinning-spacetime study reported in Ref.~\cite{majidSPINNING}
should provide a valuable starting point for such future analyses.

\section*{ACKNOWLEDGEMENTS}
This work was supported in part by a grant from the John Templeton Foundation.
MA is also supported by EU Marie Curie Actions.


\begin{thebibliography}{50}

\bibitem{dsr1} G.~Amelino-Camelia,
Int.~J.~Mod.~Phys.~{D11} (2002) 35
[gr-qc/0012051].

\bibitem{dsr2} G.~Amelino-Camelia,
Phys.~Lett.~{B510} (2001) 255
[hep-th/0012238].

\bibitem{jurekDSR1} J.~Kowalski-Glikman,
Phys.~Lett.~{A286} (2001) 391
[hep-th/0102098]

 \bibitem{leeDSRprd}
  J.~Magueijo, L.~Smolin,
%
Phys.~Rev.~{D67} (2003) 044017
[gr-qc/0207085]

\bibitem{jurekDSR2} J.Kowalski-Glikman and S.Nowak,
%
Int.~J.~Mod.~Phys.~{D12} (2003) 299
[hep-th/0204245].


\bibitem{leeDSRrainbow}
J.~Magueijo and L.~Smolin,
Class.\ Quant.\ Grav.\  {21} (2004) 1725
[gr-qc/0305055].


\bibitem{gacdsrREVIEW2010}
G. Amelino-Camelia,
Symmetry {2} (2010) 230
[arXiv:1003.3942 [gr-qc]].

\bibitem{Deser:1983tn}
  S.~Deser, R.~Jackiw and G.~'t Hooft,
  Annals Phys.\ 152 (1984) 220.

\bibitem{Deser:1988qn}
  S.~Deser and R.~Jackiw,
  Commun.\ Math.\ Phys.\ 118 (1988) 495.

\bibitem{Witten:1988hc}
  E.~Witten,
  Nucl.\ Phys.\ B311 (1988) 46.

\bibitem{Witten:1989sx}
  E.~Witten,
  Nucl.\ Phys.\ B323 (1989) 113.

\bibitem{Bais:1998yn}
  F.~A.~Bais and N.~M.~Muller,
  Nucl.\ Phys.\ B530 (1998)  349
  [hep-th/9804130].

\bibitem{Bais:2002ye}
  F.~A.~Bais, N.~M.~Muller and B.~J.~Schroers,
  Nucl.\ Phys.\ B640 (2002) 3
  [hep-th/0205021].

\bibitem{Meusburger:2003ta}
  C.~Meusburger and B.~J.~Schroers,
  Class.\ Quant.\ Grav.\ 20 (2003) 2193
  [gr-qc/0301108].

\bibitem{Meusburger:2005in}
  C.~Meusburger and B.~J.~Schroers,
  Class.\ Quant.\ Grav.\ 22 (2005) 3689
  [gr-qc/0505071].

\bibitem{Schroers:2011wn}
  B.~J.~Schroers,
  Acta Phys.\ Polon.\ Supp.\ 4 (2011) 379
  [arXiv:1105.3945 [gr-qc]].

\bibitem{Osei:2011ig}
  P.~K.~Osei and B.~JSchroers,
 J.\ Math.\ Phys.\  {53} (2012) 073510
  [arXiv:1109.4086 [gr-qc]].

\bibitem{Freidel:2005me}
  L.~Freidel and E.~R.~Livine,
  Phys.\ Rev.\ Lett.\ 96 (2006) 221301
  [hep-th/0512113].


\bibitem{Freidel:2005bb}
  L.~Freidel and E.~R.~Livine,
  Class.\ Quant.\ Grav.\ 23 (2006) 2021
  [hep-th/0502106].

\bibitem{'tHooft:1993nj}
  G.~'t Hooft,
  Class.\ Quant.\ Grav.\ 10 (1993) 1653
  [gr-qc/9305008].

\bibitem{'tHooft:1996uc}
  G.~'t Hooft,
  Class.\ Quant.\ Grav.\ 13 (1996) 1023
  [gr-qc/9601014].

\bibitem{Matschull:1997du}
  H.~-J.~Matschull and M.~Welling,
  Class.\ Quant.\ Grav.\ 15 (1998) 2981
  [gr-qc/9708054].


\bibitem{majidSPINNING}
E.~Batista and S.~Majid,
J.~Math.~Phys.~44 (2003) 107
[hep-th/0205128].


\bibitem{flaviogiuliaDESITTER}
Giulia Gubitosi, Flavio Mercati,
arXiv:1106.5710 [gr-qc].

\bibitem{goldenrule}
G. Amelino-Camelia,
Phys.\ Rev.\ D {85} (2012) 084034
[arXiv:1110.5081 [hep-th]].

\bibitem{anatomy}
G.~Amelino-Camelia, M.~Arzano, J.~Kowalski-Glikman, G.~Rosati and G.~Trevisan,
 Class.\ Quant.\ Grav.\  {29} (2012) 075007
  [arXiv:1107.1724 [hep-th]].

\bibitem{meusbschrKAPPA}
  C.~Meusburger and B.~J.~Schroers,
  Nucl.\ Phys.\ B {806} (2009) 462
  [arXiv:0805.3318 [gr-qc]].

\bibitem{majrue}
S.~Majid and H.~Ruegg,
  Phys.\ Lett.\ B {334} (1994) 348
  [hep-th/9405107].


\bibitem{lukieANNALS}
J.~Lukierski, H.~Ruegg and W.~J.~Zakrzewski,
  Annals Phys.\  {243} (1995) 90
  [hep-th/9312153].


\bibitem{prl}
  G.~Amelino-Camelia, L.~Freidel, J.~Kowalski-Glikman, L.~Smolin,
Phys.~Rev. D84 (2011) 084010
[arXiv:1101.0931  [hep-th]].

\bibitem{grf2nd}   G.~Amelino-Camelia, L.~Freidel, J.~Kowalski-Glikman, L.~Smolin,
 Gen.~Rel.~Grav. 43 (2011) 2547
 [arXiv:1106.0313 [hep-th]].

\bibitem{kodadsr}
 G.~Amelino-Camelia, L.~Smolin and A.~Starodubtsev,
 Class.~Quant.~Grav.~21 (2004) 3095
 [hep-th/0306134].

\bibitem{kodadsrJUREK}
L. Freidel, J. Kowalski-Glikman and L. Smolin,
Phys.~Rev.~D69 (2004) 044001
[hep-th/0307085].

\bibitem{oriti3D} E.R.~Livine and D.~Oriti,
 JHEP 0511 (2005) 050
 [hep-th/0509192].

\bibitem{leelaurentGRB}
 L.~Freidel, L.~Smolin,
 arXiv:1103.5626 [hep-th].

\bibitem{transverse}
G.~Amelino-Camelia, L.~Barcaroli and N.~Loret,
Int.J.Theor.Phys. 51 (2012) 3359
[arXiv:1107.3334].


\bibitem{unoEdue}
G. Amelino-Camelia and L. Smolin,
Phys. Rev. {D80} (2009) 084017
[arXiv:0906.3731].

\bibitem{sethdsr}
D.~Heyman, F.~Hinteleitner and S.~Major,
  Phys.\ Rev.\ D {69} (2004) 105016
  [gr-qc/0312089].

\bibitem{dsrphen}
G.~Amelino-Camelia, J.~Kowalski-Glikman, G.~Mandanici and A.~Procaccini,
  Int.\ J.\ Mod.\ Phys.\ A {20} (2005) 6007
  [gr-qc/0312124].

\bibitem{whataboutbob}
   G.~Amelino-Camelia, M.~Matassa, F.~Mercati and G.~Rosati,
  Phys.\ Rev.\ Lett.\  {106} (2011) 071301
  [arXiv:1006.2126 [gr-qc]].

\bibitem{cortesRL}
J.~M.~Carmona, J.~L.~Cortes, D.~Mazon and F.~Mercati,
  Phys.\ Rev.\ D { 84} (2011) 085010
  [arXiv:1107.0939 [hep-th]].

\bibitem{fuzzy1} G.~Amelino-Camelia, V.~Astuti and G.~Rosati,
arXiv:1206.3805.

\end{thebibliography}
\end{document}